\documentclass[letterpaper, 10 pt, conference]{ieeeconf}  

\IEEEoverridecommandlockouts                              

\usepackage{amsmath} 
\usepackage{amssymb}  
\usepackage{color}
\usepackage{graphicx}

\def\be{\begin{equation}}
\def\ee{\end{equation}}
\def\ba{\begin{array}}
\def\ea{\end{array}}

\newtheorem{example}{Example}
\newtheorem{rem}{Remark}
\newtheorem{assum}{Assumption}
\newtheorem{definition}{Definition}
\newtheorem{lem}{Lemma}
\newtheorem{prop}{Proposition}

\newtheorem{cor}{Corollary}

\newcommand{\smat}[1]{\ensuremath{\left[\begin{smallmatrix}#1\end{smallmatrix}\right]}}

\def\qedp{\hspace*{\fill}~{\tiny $\blacksquare$}} 
\def\qeds{\hspace*{\fill}~{\tiny $\Box$}}

\title{\LARGE \bf
Data-driven control of nonlinear systems from input-output data*
}

\author{X. Dai$^{1}$ and C. De Persis$^{1}$ and N. Monshizadeh$^{1}$ and P. Tesi$^{2}$
\thanks{*This publication is also part
of the project Digital Twin with project number P18-03 of
the research programme TTW Perspective which is (partly)
financed by the Dutch Research Council (NWO).}
\thanks{$^{1}$X. Dai,
C. De Persis and N. Monshizadeh are with Institute of Engineering and Technology Groningen, University of Groningen, 9747 AG Groningen, 
The Netherlands. Email: {\tt\small \{x.dai,c.de.persis,n.monshizadeh\}@rug.nl}.
}%
\thanks{$^{2}$P. Tesi is with the Department of Information Engineering, 
University of Florence, 50139 Florence, Italy. Email: {\tt\small pietro.tesi@unifi.it}. 
}%
}

\begin{document}

\maketitle
\thispagestyle{empty}
\pagestyle{empty}

\begin{abstract}
The design of controllers from data for nonlinear systems is a challenging problem. In a recent paper, 
De Persis, Rotulo and Tesi, ``Learning controllers from data via approximate nonlinearity cancellation," IEEE Transactions on Automatic Control, 2023,
a method to learn controllers that make the closed-loop system stable and dominantly linear was proposed. The approach leads to a simple solution based on data-dependent semidefinite programs. The method uses input-state measurements as data, while in a realistic setup it is more likely that only input-{\em output} measurements are available. In this note we report how the design principle of 
the above mentioned paper 
can be adjusted to deal with input-output data and obtain dynamic output feedback controllers in a favourable setting. 
\end{abstract}

\section{INTRODUCTION}

Learning controllers from data is of uttermost importance and a fascinating topic, with foundations in both control theory and data science. Several recent approaches have been proposed for data-driven control, initially focusing, as is natural, on linear systems, e.g. \cite{JC-JL-FD:18, de2019formulas, 
van2020data, berberich2020robust}. 
For nonlinear systems, some results have appeared as well, mostly focusing on special classes of nonlinear systems, bilinear  \cite{bisoffi2020data,yuan2021data}, polynomial \cite{guoTAC2021poly,Nejati2022poly,luppi2021data}, rational \cite{strasser2021data} or with quadratic nonlinearities \cite{luppi2022data}, \cite{Cheah2022NL}. Other approaches consist of approximating  general nonlinear control systems to classes for which data-driven design is possible \cite{Guo2022Taylor,Martin2022Taylor} or expressing nonlinear systems via a dictionary of known functions, in which case the design can aim at making the closed-loop system dominantly linear 
\cite{2021cancellation}
or prescribing a desired output signal 
\cite{alsalti2023data}. 

The understanding of the topic is far from having reached a mature phase, even in the case full measurements of the state are available. Yet, it can be argued that the use of these data-dependent design schemes in practice very much rely on the possibility that they work with output measurements data only, which dispenses the designer from requiring to know the state of the system -- a very restrictive prior in many cases. In this paper we report on some early results on using data-driven control techniques in conjunction with input/output data for discrete-time nonlinear systems.  

{\it Related work.}  Even when a model is known, output feedback control for nonlinear systems is a challenging open problem \cite[Section 8.7]{bernard2022observers}. The certainty equivalence principle, which is valid for linear systems, is hard to extend to a nonlinear setting. Nonetheless, certain nonlinear discrete-time versions of the certainty equivalence principle have been obtained \cite{kazakos1993stabilization}. 
In \cite{messina2005discrete},  the state in a globally stabilizing state feedback (possibly generated by a finite horizon model predictive scheme) is replaced by an estimate provided by an observer under a uniform observability assumption to obtain a globally stabilizing output feedback controller.   

The important uniform observability property \cite{gauthier1981observability,grizzle1990newton,hanba2009uniform} can be explored in different ways in the context of learning control from data. Since it  guarantees the existence of an injective map from input/output sequences to the state, deep neural networks can be trained to approximate such 
a map and provide estimates of the state to be used in the given input-to-state stabilizing feedback, obtaining a locally asymptotically stable closed-loop system \cite{marchi2021safety}. The injective map can also be used to define the regression relating the input/output sequences of the system and deep neural networks can be used to learn such a regression \cite{janny2022learning}. However, to the best of our knowledge there are very few other attempts at designing controllers for nonlinear system from input/output data.
\par 
{\it Contribution.} The aim of this note is to start the investigation of feedback design from input/{\em output} data for nonlinear discrete-time systems. We adopt the notion of uniform observability, which allows us to extend some of the design procedures introduced in \cite{de2019formulas}. Namely, we consider past inputs and outputs as fictitious state variables and obtain a form of the system for which the data-driven ``state" feedback design techniques for nonlinear systems of \cite{2021cancellation} can be used. 
The implementation of the controller is then carried out by replacing the past input/output measurements with the quantities returned by a dead-beat observer of the output and a chain of integrators driven by the input. A formal analysis of the stability of the overall closed-loop system is then presented along with a discussion about the proposed solution. 
\par
In Section \ref{sec:preliminaries} we recall the notion of observability that we adopt for our analysis and introduce an auxiliary system that reproduces the input/output behaviour of the system to control. The auxiliary system is extended  in Section \ref{sec:dyn-ext}  with a chain of integrators that provides the past inputs of the system to be used in the controller. The design of the output feedback dynamic controller based on input/output data is presented in Section \ref{sec:control-design}. The analysis of the closed-loop system to show the convergence of the system's and the controller's state to the origin is the topic of Section \ref{sec:convergence-true-state}, along with a discussion  of the result.

\section{Preliminaries}\label{sec:preliminaries}
We consider the single-input single-output nonlinear discrete-time system
\be\label{nonl}
\ba{rl}
x^+= &  f(x,u)\\
y = & h(x)
\ea
\ee
where $x\in \mathbb{R}^n$, $u,y\in \mathbb{R}$, $f(0,0)=0$ and $h(0)=0$. $f,h$ are continuous functions of their arguments with domains $\mathbb{R}^n\times \mathbb{R}$ and $\mathbb{R}^n$. These functions are unknown. The dimension of the state-space $n$ is not necessarily known.

\subsection{Dataset}

A dataset consisting of open-loop input-output measurements 
\be\label{dataset}
\mathcal{D}:=\{(u(k),y(k))\}_{k=0}^{N+T-1}
\ee
is available, where the positive integers $N,T$ will be specified later. The samples in the dataset are obtained from off-line experiment(s) conducted on system \eqref{nonl}, hence they satisfy the equations \eqref{nonl}, namely 
\[
\ba{rl}
x(k+1)= &  f(x(k),u(k))\\
y(k) = & h(x(k)),\quad \forall k=0,1, \ldots,  N+T-1
\ea
\]
For our purpose of designing an output feedback controller from $\mathcal{D}$ it is not required that all the samples of the dataset are sequentially obtained in a single experiment. In fact, even multiple experiments collecting $N+T$ samples suffice. This is useful especially when dealing with unstable dynamics. 

\subsection{Uniform Observability}

The problem of interest is to design an output feedback controller that stabilizes the nonlinear system, based on the dataset  $\mathcal{D}$. To this purpose, we need to infer the behavior of the state $x$ from input-output measurements, for which suitable ``observability" conditions on the system \eqref{nonl} are required. Before stating them, we introduce some notation. We let 
\be\label{F}\ba{rll}
F^0(x) := & x\\
F^1(x,v_0):= & f(x,v_0)\\
F^{k+1}(x,v_0, \ldots, v_k):= & f(F^k(x,v_0, \ldots, v_{k-1}),v_k),& \!\!\!\!\!k\ge 1
\ea
\ee 
Note that \eqref{F} gives $x(k)=F^N(x(k-N),u_{[k-N,k-1]})$.
To reduce the notational complexity, we introduce $v_{[0,k]}$, which denotes the sequence of values $v_0, \ldots, v_k$.
Hence, the last identity above is rewritten as $F^{k+1}(x,v_{[0,k]}):= f(F^k(x,v_{[0,k-1]}),v_k)$. In what follows, we will  use symbols like $v_{[0,k]}$ also to denote the vector $\smat{v_0 & v_1 & \ldots & v_k}^\top$.
\medskip
\par
The following is the main assumption on system \eqref{nonl}.
\smallskip

\begin{assum}\label{asspt-uo-on-set}
Let $\mathcal{X}\subset \mathbb{R}^n$ and $\mathcal{U}\subset \mathbb{R}$ be compact sets such that $\mathcal{X}\times \mathcal{U}$ contains the origin of $\mathbb{R}^{n+1}$. There exists $N\in \mathbb{Z}_{>0}$ such that, for any
$v_{[0,N-2]}\in  \mathcal{U}^{N-1}$, 
the mapping 
\be\label{Phi_N}
\Phi_N(x, v_{[0,N-2]})=
\begin{bmatrix}
h\circ F^0(x)\\
h\circ F^1(x,v_0)\\
\vdots \\
h\circ F^{N-1}(x,v_{[0,N-2]})
\end{bmatrix}
\ee
is injective as a function of $x$ on $\mathcal{X}$. \qeds
\end{assum}

\medskip

Following  \cite[Definition 1]{hanba2009uniform},
we refer to the assumption above as a uniform observability on $\mathcal{X}$ property. It is observed in \cite{hanba2009uniform} that, if $f,h$ are continuously differentiable functions,  uniform observability is not restrictive in the sense that a nonuniform distinguishability property and a nonuniform observability rank condition imply uniform observability. Since for any $M\ge N$ the mapping  $\Phi_M$ remains injective, 
we do not need to know the smallest $N$ for which Assumption \ref{asspt-uo-on-set} holds.
\par
For any
$v_{[0,N-2]}\in  \mathcal{U}^{N-1}$, the function 
\[
\Phi_N(\cdot, v_{[0,N-2]})\colon \mathcal{X}\to \mathbb{R}^N
\]
such that $x\mapsto w= \Phi_N(x, v_{[0,N-2]})$, is injective on  $\mathcal{X}$ and one can define a left inverse 
\[
\Psi_N(\cdot,v_{[0,N-2]})\colon \Phi_N(\mathcal{X}, v_{[0,N-2]})\to \mathbb{R}^n
\]
such that $\Psi_N(\Phi_N(x, v_{[0,N-2]}),v_{[0,N-2]})=x$ for all $x\in \mathcal{X}$.

\subsection{An auxiliary system}

We introduce a system equivalent to \eqref{nonl} which is better suited for control design. By equivalent it is meant that the new system has the same input-output behavior of system \eqref{nonl} when properly initialized. We use this auxiliary system for control design purposes. Later on we show the effect of the designed controller on the actual system \eqref{nonl}.

For any $v_{[0, N-1]}\in \mathbb{R}^N$, define the functions 
\be\label{mappings}
\ba{rl}
\psi(w, v_{[0, N-1]}): = &F^N(\Psi_N(w,v_{[0, N-2]}), v_{[0, N-1]})\\
\tilde h(w,v_{[0, N-1]}):= & h\circ \psi(w, v_{[0, N-1]}) \\[0.5mm]
\tilde f (w,v_{[0, N-1]}) :=& 
A_c w+B_c \tilde h(w, v_{[0, N-1]})
\ea\ee
with the pair $(A_c, B_c)\in \mathbb{R}^{N\times N}\times 
\mathbb{R}^{N}$ in the Brunovsky form. 
The domain of $\psi(\cdot, v_{[0, N-1]}), \tilde h(\cdot, v_{[0, N-1]})$, $\tilde f(\cdot, v_{[0, N-1]})$ is  
$\Phi_N(\mathcal{X}, v_{[0,N-2]})$. 
Under the standing assumptions on $f,h$, 
these functions are continuous and zero at $(w,v)=(0,0)$.

\smallskip

In the result below, for a  $k\in \mathbb{Z}$, we let 
$u_{[k-N, k-1]}$ be
an input sequence applied to system \eqref{nonl} and 
$y_{[k-N, k-1]}$ 
its output response from some initial condition $x(k-N)$.

\begin{lem}\label{lem-equiv}
Let system \eqref{nonl} satisfy Assumption \ref{asspt-uo-on-set}. Consider arbitrary $k_0\in \mathbb{Z}$, 
$x(k-N)\in \mathcal{X}$ and $u_{[k-N, k-1]}\in \mathcal{U}^N$  for all $k\in \mathbb{Z}_{\ge k_0}$. 
Consider the system
\be\label{nonl-eq}
\ba{rl}
w^+= &
\tilde f (w,v)\\
y_w = &  \tilde h(w,v)
\ea\ee
with $\tilde f, \tilde h$ defined in \eqref{mappings}. 
If
the input $v(k)$ applied to \eqref{nonl-eq} satisfies  
$v(k)=u_{[k-N,k-1]}$ 
for all $k\in \mathbb{Z}_{\ge k_0}$ and the initial condition of \eqref{nonl-eq} is set to  
$w(k_0)=y_{[k_0-N, k_0-1]}$, 
then 
\[
w(k)=y_{[k-N, k-1]}, \quad 
y_w(k) = y(k), \quad \forall k\in \mathbb{Z}_{\ge k_0}.  
\]
Furthermore, $x(k)= \psi(w(k),v(k))$, for all $k\in \mathbb{Z}_{\ge k_0}$. \qeds
\end{lem}

\medskip

{\it Proof.} For the sake of completeness, it is given in Appendix \ref{app:proof-lemma-1}. \qedp

\medskip

\begin{example} \label{example} We consider \cite[Example 5]{2021cancellation}
\begin{subequations} 
\label{eq:sys_example_pendulum}
\begin{alignat}{2}
& x_1^+ = x_1 + T_s x_2 \\   
& x_2^+ = \displaystyle  \frac{T_s  g}{\ell} \sin x_1 + \left( 1 - \frac{T_s  \mu}{m \ell^2} \right) x_2 +
\frac{T_s }{m \ell } (\cos x_1)u \,,
\end{alignat}
\end{subequations}
with $y=x_1$. 
We compute 
\[
\Phi_2(x,v)=
\begin{bmatrix}
x_1\\
x_1 + T_s x_2
\end{bmatrix}
\] 
which is globally invertible (Assumption \ref{asspt-uo-on-set} holds with $N=2$.) with 
\[
\Psi_2(w,v)=
\begin{bmatrix}
w_1\\
\frac{w_2-w_1}{T_s}
\end{bmatrix}
\]
Hence $\psi(w, v_0,v_1)={\rm col}(\psi_1(w, v_0,v_1), \psi_2(w, v_0,v_1))$, where
\be\label{psi}\ba{rl}
\psi_1(w, v_0,v_1)= &
w_2+\frac{T_s^2 g}{\ell}\sin w_1+ \left(1-\frac{T_s\mu }{m\ell^2}\right)(w_2-w_1)\\[2mm]
&\hspace{3.5cm}+\frac{T_s^2 }{m\ell}(\cos w_1) v_0\\[2mm]
\psi_2(w, v_0,v_1)= & \frac{T_s g}{\ell}\sin w_2
+ \left(1-\frac{T_s\mu }{m\ell^2}\right)
\left(
\frac{T_s g}{\ell}\sin w_1 
\right.\\
&\hspace{-2cm}
\left.
+ 
\left(1-\frac{T_s\mu }{m\ell^2}\right) \frac{w_2-w_1}{T_s}
+\frac{T_s }{m\ell}(\cos w_1) v_0
\right)
+\frac{T_s }{m\ell}(\cos w_2) v_1
\ea\ee
From which, one computes
\[\ba{l}
\tilde h(w,v_0)= 
w_2+\frac{T_s^2 g}{\ell}\sin w_1+ \left(1-\frac{T_s\mu }{m\ell^2}\right)(w_2-w_1)\\
\hspace{3.5cm}+\frac{T_s^2 }{m\ell}(\cos w_1) v_0\\
 = 
\left(-1+\frac{T_s\mu }{m\ell^2}\right) w_1+ \left(2-\frac{T_s\mu }{m\ell^2}\right)w_2
+\frac{T_s^2 g}{\ell}\sin w_1\\
\hspace{4cm}+\frac{T_s^2 }{m\ell}(\cos w_1) v_0
\ea
\]
Hence the equivalent representation is given by 
\[
\ba{l}
w^+=  \begin{bmatrix}
w_2\\
\tilde h(w,v_0)\end{bmatrix}, \quad 
y_w= 
\tilde h(w,v_0)
\ea
\]
The original state $x$ is obtainable from the solution of the system above via the expression
\[
x=\psi(w, v_0,v_1)
\]
where $\psi$ is as in \eqref{psi}. 
For this example $\mathcal{X}=\mathbb{R}^2$ and $\mathcal{U}=\mathbb{R}$. \qedp
\end{example}

\section{Design of an output feedback controller from data}\label{sec:control-design}

\subsection{A dynamic extension}\label{sec:dyn-ext}

System \eqref{nonl-eq}  
is driven by the past $N$ samples of $u$, which is the input to \eqref{nonl}. These past values  are obtained by adding a chain of integrators to the dynamics \eqref{nonl-eq}

\be\label{chain-integrators}
\ba{rl}
\xi^+ = & A_c\,\xi +B_c u
\ea
\ee
with the interconnection condition 
\[
v=  \xi
\]
which returns the 
system 
\be\label{nonl-eq-extended}
\ba{rl}
w^+= & \tilde f (w,\xi)\\
\xi^+ = & A_c\,\xi +B_c u\\
y= & \tilde h (w,\xi)
\ea
\ee
Once the system's state satisfies $(w(\overline k), \xi(\overline k))=(y_{[\overline k-n,\overline k-1]}, u_{[\overline k-n,\overline k-1]})$ for some $\overline k\in \mathbb{Z}$, the input-output behavior of this system matches the one of \eqref{nonl} for all $k\ge \overline k$. We will discuss later on the availability of such initial condition at a time $\overline k$.

\subsection{Control input design}
To obtain $u$ that drives the chain of integrators making the dynamic controller,  we argue as in \cite{de2019formulas, 2021cancellation}. We first introduce the following:
\begin{assum}\label{asspt-dictionary}
For any $\xi\in \mathcal{U}^{N}$ and any $w\in \Phi_N(\mathcal{X}, \xi_{[1, N-1]})$, where $\xi_{[1, N-1]}$ denotes the first $N-1$ entries of $\xi$, it holds that 
$\tilde h(w, \xi)=\alpha Z(w,  \xi)$, where  
$Z(w,\xi)\in \mathbb{R}^{S}$ is a  vector of known continuous functions and $\alpha\in \mathbb{R}^{1\times S}$ is an unknown vector. \qeds
\end{assum}
\par
This is a technical assumption due to the need to give the nonlinearities of \eqref{nonl-eq-extended} a form for which the controller design is possible.  Although it is restrictive,
\cite[Section VI.B]{2021cancellation} 
bypasses such an assumption 
by expressing $\tilde h(w, \xi)$ as $\alpha Z(w,  \xi)+d(w,\xi)$, where the term $d(w,\xi)$ represents the nonlinearities that were excluded from $Z(w,  \xi)$, and then analyzing the stability of the system in the presence of the neglected nonlinearity $d(w,\xi)$. This analysis goes beyond the scope of this paper. 
\par
We consider the case in which the function $Z(w,  \xi)$ comprises both a linear part and a nonlinear part $Q(w,  \xi)$, i.e.
\[
Z(w,  \xi) = \begin{bmatrix}
w\\
\xi\\
Q(w,  \xi)
\end{bmatrix}
\]
The system \eqref{nonl-eq-extended} can then be written as 
\be\label{nonl-eq-extended-closed-loop}
\ba{rl}
\begin{bmatrix}
w^+\\
\xi^+
\end{bmatrix}
=&
A \begin{bmatrix}
w\\
\xi
\end{bmatrix}
+
B_1
u
+
B_2\alpha Z(w,\xi)\\
y= & \alpha Z (w,\xi)
\ea
\ee
where
\[
A:=
\begin{bmatrix}
A_c & 0 \\
0 & A_c
\end{bmatrix},\;
B_1:=
\begin{bmatrix}
0 \\
B_c
\end{bmatrix},\;
B_2:=
\begin{bmatrix}
B_c \\
0
\end{bmatrix}
\]
and the pair $(A_c, B_c)$ is in the Brunovsky canonical form. 

\par
We focus on the case in which the input $u$ is designed as a function of $Z(w,  \xi)$, i.e.
\be\label{control}
u=\kappa Z(w,  \xi)
\ee
where $\kappa \in \mathbb{R}^{1\times S}$ is the control gain. 
Write the closed-loop system \eqref{nonl-eq-extended-closed-loop}-\eqref{control} as 
\be\label{nonl-eq-extended-closed-loop-actual}
\ba{rl}
\begin{bmatrix}
w^+\\
\xi^+
\end{bmatrix}
=&
A \begin{bmatrix}
w\\
\xi
\end{bmatrix}
+
B_1
\kappa Z(w,\xi)
+
B_2\alpha Z(w,\xi)\\
y= & \alpha Z (w,\xi)
\ea
\ee
The system is defined for any $\xi\in \mathcal{U}^{N}$ and any $w\in \Phi_N(\mathcal{X}, \xi_{[1, N-1]})$. 

\subsection{Data-dependent representation of the closed-loop system}
Preliminary to the design of the controller is a data-dependent representation of the closed-loop system. We first introduce some notation. Recall the dataset in \eqref{dataset} and introduce, for  $i=0,\ldots, T-1$,
\[
U(i):=\begin{bmatrix}
u(i) \\
u(i+1)\\
\vdots\\
u(i+N-1)
\end{bmatrix},
Y(i):=\begin{bmatrix}
y(i) \\
y(i+1)\\
\vdots\\
y(i+N-1)
\end{bmatrix}
\]
We assume that the samples of the dataset evolve in the domain of definition of \eqref{nonl-eq-extended-closed-loop-actual}.

\begin{assum}\label{asspt:dataset-in-the-right-domain}
For any $i=0,\ldots, T-1$, $U(i)\in \mathcal{U}^{N}$ and 
$Y(i)\in \Phi_N(\mathcal{X}, u_{[i,i+N-2]})$. \qeds
\end{assum}

\smallskip

We let:
\be\label{dataset-matrices}
\ba{rl}
Y_0 & := 
\begin{bmatrix}
Y(0) & Y(1) & \ldots &Y(T-1)
\end{bmatrix}
\\[2mm]
V_0& :=
\begin{bmatrix}
U(0) & U(1) & \ldots & U(T-1)
\end{bmatrix}
\\[2mm]
Y_1&:= 
\begin{bmatrix}
Y(1) & Y(2) & \ldots &Y(T)
\end{bmatrix}
\\[2mm]
V_1 & := 
\begin{bmatrix}
U(1) & U(2) & \ldots & U(T)
\end{bmatrix}
\\[2mm]
Q_0& := 
\begin{bmatrix}
Q(0) & Q(1) & \ldots & Q(T-1)
\end{bmatrix}
\\[2mm]
U_0 & :=
\begin{bmatrix}
u(N) & u(N+1) & \ldots u(N+T-1)
\end{bmatrix}
\ea
\ee
In the definition of $Q_0$, we are using the shorthand notation 
$Q(i)$ for $Q(Y(i), U(i))$. 
Under Assumption   \ref{asspt:dataset-in-the-right-domain}, bearing in mind the dynamics \eqref{nonl-eq-extended-closed-loop}, the dataset-dependent matrices introduced in \eqref{dataset-matrices} satisfy
\be\label{dataset-matrices-identity}
\begin{bmatrix}
Y_1\\
V_1
\end{bmatrix}
=A \begin{bmatrix}
Y_0\\
V_0
\end{bmatrix}
+
B_1 
U_0+
B_2\alpha \begin{bmatrix}
Y_0\\
V_0\\
Q_0
\end{bmatrix}
\ee

\begin{rem} \label{rem.mult-experiments}
{\em (Multiple experiments)} This identity is obtained from the $T$ identities 
\[
\ba{l}
\begin{bmatrix}
Y(i+1)\\
U(i+1)
\end{bmatrix}
=A \begin{bmatrix}
Y(i)\\
U(i)
\end{bmatrix}
+
B_1 
u(i+N)+
B_2\alpha \begin{bmatrix}
Y(i)\\
U(i)\\
Q(i)
\end{bmatrix},\\
\hspace{5cm}i=0,\ldots, T-1
\ea
\]
We note that, for each $i$, the identity does not require the quantities $Y(i), U(i),  Y(i+1), U(i+1), u(i)$ to be related to the corresponding quantities for $i+1$. In other words, we could run $T$ $N$-long independent experiments and collect the resulting input-output samples in 
\[\ba{l}
Y_0^j:=\begin{bmatrix}
y^j(0) \\
y^j(1)\\
\vdots\\
y^j(N-1)
\end{bmatrix},
\; 
U_0^j:=\begin{bmatrix}
u^j(0) \\
u^j(1)\\
\vdots\\
u^j(N-1)
\end{bmatrix},
\\
Y_1^j:=\begin{bmatrix}
y^j(1) \\
y^j(2)\\
\vdots\\
y^j(N)
\end{bmatrix},
\; 
U_1^j:=\begin{bmatrix}
u^j(1) \\
u^j(2)\\
\vdots\\
u^j(N)
\end{bmatrix}
\ea\]
where $j=0,\ldots, T-1$ denotes the number of the experiment, and $\{u^j(k), y^j(k)\}_{k=0}^{N}$
are the input-output samples of the experiment $j$. We could then redefine the matrices in \eqref{dataset-matrices} as 
\[
\ba{rl}
Y_0 & := 
\begin{bmatrix}
Y_0^0 & Y_0^1 & \ldots &Y_0^{T-1}
\end{bmatrix}
\\[2mm]
V_0& :=
\begin{bmatrix}
U_0^0 & U_0^1 & \ldots & U_0^{T-1}
\end{bmatrix}
\\[2mm]
Y_1&:= 
\begin{bmatrix}
Y_1^0 & Y_1^1 & \ldots &Y_1^{T-1}
\end{bmatrix}
\\[2mm]
V_1 & := 
\begin{bmatrix}
U_1^0 & U_1^1 & \ldots & U_1^{T-1}
\end{bmatrix}
\\[2mm]
Q_0& := 
\begin{bmatrix}
Q_0^0& Q_0^1&  \ldots & Q_0^{T-1}
\end{bmatrix}
\\[2mm]
U_0 & :=
\begin{bmatrix}
u^0(N) & u^1(N) & \ldots u^{T-1}(N)
\end{bmatrix}
\\[2mm]
\ea
\]
and the identity \eqref{dataset-matrices-identity} would still apply. \qedp
\end{rem}

We establish the following:
\begin{lem}\label{data-dependent-outp-feedback-repr}
Let Assumptions \ref{asspt-uo-on-set}, \ref{asspt-dictionary} and \ref{asspt:dataset-in-the-right-domain} hold. Consider any matrices $\kappa\in\mathbb{R}^{1\times S}, G\in\mathbb{R}^{T\times S}$ that satisfy the relation
\be\label{rich-data}
\begin{bmatrix}
\kappa\\
\hline
I_S 
\end{bmatrix}
= 
\begin{bmatrix}
U_0\\
\hline
Y_0\\
V_0\\
Q_0
\end{bmatrix}
G
\ee
and partition $G$ as 
\[
G=
\begin{bmatrix}
G_1 & G_2 
\end{bmatrix}
\]
where $G_1\in \mathbb{R}^{T\times 2N}, G_2\in \mathbb{R}^{T\times (S-2N)}$.
Then the 
closed-loop system 
\eqref{nonl-eq-extended-closed-loop-actual}
can be  written as
\[
\begin{bmatrix}
w^+\\
\xi^+
\end{bmatrix}
=M \begin{bmatrix}
w\\
\xi
\end{bmatrix}+N Q(w,\xi)
\]
where 
\be\label{M-N}
M=  
\mathcal{X}_1 G_1, \; 
N=
\mathcal{X}_1 G_2,\; 
\mathcal{X}_1 =  \begin{bmatrix}
Y_1\\
V_1
\end{bmatrix}.  
\ee
\qeds
\end{lem}

{\it Proof.} 
For any $\xi\in \mathcal{U}^{N}$ and any $w\in \Phi_N(\mathcal{X}, \xi_{[1, N-1]})$, it holds
\[\ba{rl}
&
A \begin{bmatrix}
w\\
\xi
\end{bmatrix}
+
B_1
\kappa Z(w,\xi)
+
B_2\alpha Z(w,\xi)\\[3mm]
\stackrel{\eqref{rich-data}}{=}&
A \begin{bmatrix}
w\\
\xi
\end{bmatrix}
+
B_1
U_0 G_1 \begin{bmatrix}
w\\
\xi
\end{bmatrix}
+
B_1
U_0 G_2 Q(w,\xi)
\\
&
\hspace{4cm}
+
B_2\alpha
\begin{bmatrix}
Y_0\\
V_0\\
Q_0
\end{bmatrix}
G
Z(w,\xi)
\\
\stackrel{\eqref{dataset-matrices-identity}}{=} & 
A \begin{bmatrix}
w\\
\xi
\end{bmatrix}
+
B_1
U_0 G_1 \begin{bmatrix}
w\\
\xi
\end{bmatrix}
+
B_1
U_0 G_2 Q(w,\xi)
\\[3mm]
&\hspace{2cm}
+
\left(
\begin{bmatrix}
Y_1\\
V_1
\end{bmatrix}
-A \begin{bmatrix}
Y_0\\
V_0
\end{bmatrix}
-
B_1
U_0\right)G
Z(w,\xi)\\
=& 
\left(
A + 
\left(
\begin{bmatrix}
Y_1\\
V_1
\end{bmatrix}
-A \begin{bmatrix}
Y_0\\
V_0
\end{bmatrix}
\right)G_1 
\right)
\begin{bmatrix}
w\\
\xi
\end{bmatrix}
\\
&\hspace{3cm}
+
\left(
\begin{bmatrix}
Y_1\\
V_1
\end{bmatrix}
-A \begin{bmatrix}
Y_0\\
V_0
\end{bmatrix}
\right)G_2
Q(w,\xi).
\ea\]
By \eqref{rich-data}, we obtain that
\[
\begin{bmatrix}
Y_0\\
V_0\\
\end{bmatrix}
\begin{bmatrix}
G_1 
& 
G_2
\end{bmatrix}
=
\begin{bmatrix}
I_{2N}
& 
0_{S-2N}
\end{bmatrix}
\]
Hence,
\[\ba{rl}
A + 
\left(
\begin{bmatrix}
Y_1\\
V_1
\end{bmatrix}
-A \begin{bmatrix}
Y_0\\
V_0
\end{bmatrix}
\right)G_1 = &\mathcal{X}_1 G_1\\[4mm]
\left(\begin{bmatrix}
Y_1\\
V_1
\end{bmatrix}
-A \begin{bmatrix}
Y_0\\
V_0
\end{bmatrix}
\right)G_2
=& \mathcal{X}_1 G_2.
\ea\]
\qedp
\par
Let the set of real-valued symmetric matrices of dimension $n \times n$ be denoted by $\mathbb{S}^{n \times n}$. 
This data-dependent representation leads to the following local stabilization result:
\begin{prop}\label{prop:approx}
Let Assumptions \ref{asspt-uo-on-set}, \ref{asspt-dictionary} and \ref{asspt:dataset-in-the-right-domain} hold. 
Consider  the following SDP in the decision variables $\mathcal{P}_1 \in \mathbb S^{2N \times 2N}$,
$\mathcal{Y}_1 \in \mathbb R^{T \times 2 N}$, and
$G_2 \in \mathbb R^{T \times (S-2N)}$:
\begin{subequations}
\label{eq:2SDP}
\begin{alignat}{2}
\textrm{minimize}_{\mathcal{P}_1,\mathcal{Y}_1,G_2} \quad
& \|\mathcal{X}_1 G_2\| \label{eq:2SDPa} \\ 
\textrm{subject to} \quad 
& \begin{bmatrix}
Y_0\\
V_0\\
Q_0
\end{bmatrix}
 \mathcal{Y}_1 = \begin{bmatrix}  \mathcal{P}_1 \\ 0_{(S-2N) \times 2N} \end{bmatrix} \,,
\label{eq:2SDP1} \\
& 
\begin{bmatrix}  \mathcal{P}_1 & (\mathcal{X}_1  \mathcal{Y}_1)^\top  \\ 
\mathcal{X}_1  \mathcal{Y}_1 & \mathcal{P}_1 \end{bmatrix}  \succ 0 \,, 
\label{eq:2SDP2} \\
& \begin{bmatrix}
Y_0\\
V_0\\
Q_0
\end{bmatrix} G_2 = \begin{bmatrix} 0_{2N \times (S-2N)} \\ I_{S-2N} \end{bmatrix} \,.
\label{eq:2SDP4} 
\end{alignat}
\end{subequations}
Assume that 
\begin{equation} \label{eq:limitQx}
\lim_{|(w,\xi)|\to 0} \frac{|Q(w,\xi)|}{|(w,\xi)|}=0 \,.
\end{equation}
If the SDP is feasible then 
\be\label{chain-integrators-with-u}
\xi^+ = A_c \xi +B_c u 
\ee
with 
\be\label{u=kZ}
u= \kappa Z(w,\xi)
\ee
and $\kappa$ as in
\begin{eqnarray} \label{eq:kappa_SDP}
\kappa = U_0 \begin{bmatrix} \mathcal{Y}_1 & G_2 \end{bmatrix}
\begin{bmatrix} \mathcal{P}_1^{-1}  & 
0_{2N \times (S-2N)} \\
0_{(S-2N) \times 2N} & I_{S-2N}
\end{bmatrix}
\end{eqnarray}
renders the origin $(\overline w, \overline \xi)=(0,0)$ an asymptotically stable equilibrium of 
\be\label{nonl-eq-extended-actual-closed-loop}
\ba{rl}
w^+= & \tilde f (w,\xi)\\
\xi^+ =& A_c \xi +B_c \kappa Z(w,\xi) \\
y= & \tilde h (w,\xi).
\ea
\ee
\qeds
\end{prop}

{\it Proof.} Set $G_1=\mathcal{Y}_1\mathcal{P}_1^{-1}$. Then \eqref{eq:2SDP1}, \eqref{eq:2SDP4}
imply
\[
I_S 
= 
\begin{bmatrix}
Y_0\\
V_0\\
Q_0
\end{bmatrix}
\begin{bmatrix}
G_1 & G_2
\end{bmatrix}
\]
which along with the definition of $\kappa$ in \eqref{eq:kappa_SDP}, namely 
$\kappa = U_0 \smat{G_1 & G_2}$, implies \eqref{rich-data}. Hence, the data-dependent representation of system \eqref{nonl-eq-extended-closed-loop-actual} given in   Lemma \ref{data-dependent-outp-feedback-repr} holds. By Schur complement, the constraint \eqref{eq:2SDP2} is equivalent to 
\[
\mathcal{P}_1 -(\mathcal{X}_1  \mathcal{Y}_1)^\top \mathcal{P}_1^{-1}
(\mathcal{X}_1  \mathcal{Y}_1) \succ 0 \,, 
\]
Pre- and post-multiplying by $\mathcal{P}_1^{-1}$ and bearing in mind the definition of $G_1$ we obtain 
\[
\mathcal{P}_1^{-1} -(\mathcal{X}_1 G_1)^\top \mathcal{P}_1^{-1}
(\mathcal{X}_1 G_1) \succ 0 \,, 
\]
which shows that $V(w,\xi)= \begin{bmatrix} w^\top & \xi^\top\end{bmatrix} \mathcal{P}_1^{-1} 
\begin{bmatrix} w \\ \xi \end{bmatrix}$ is a Lyapunov function for the linear part of the closed-loop system. In particular note that the domain of definition of the function $V(w,\xi)$ is the same as the one of system \eqref{nonl-eq-extended-actual-closed-loop}, hence, $V(w,\xi)$ is defined at the origin. We have
\[\ba{rl}
& V(w^+,\xi^+)-V(w,\xi)\\
=& \begin{bmatrix} w^\top & \xi^\top\end{bmatrix}
((\mathcal{X}_1 G_1)^\top \mathcal{P}_1^{-1}
(\mathcal{X}_1 G_1)- \mathcal{P}_1^{-1})\begin{bmatrix} w \\ \xi \end{bmatrix}\\
&+2  \begin{bmatrix} w^\top & \xi^\top\end{bmatrix}
(\mathcal{X}_1 G_1)^\top \mathcal{P}_1^{-1} \mathcal{X}_1 G_2 Q(w,\xi)\\
&+Q(w,\xi)^\top (\mathcal{X}_1 G_2)^\top\mathcal{P}_1^{-1} \mathcal{X}_1 G_2 Q(w,\xi)
\ea
\]
In view of \eqref{eq:limitQx}, $V(w^+,\xi^+)-V(w,\xi)<0$ in a neighborhood of the origin. This shows the claim. 
\qedp

\subsection{Region of Attraction} 
 Proposition \ref{prop:approx} provides a local stabilization result. Following \cite{2021cancellation}, Proposition \ref{prop:approx} can be extended to provide an estimate of the Region of Attraction  (ROA) of the system \eqref{nonl-eq-extended-actual-closed-loop}. 
First we recall the following definitions.
\begin{definition}
\cite[Definition 13.2]{haddad2011nonlinear}
Suppose that $\overline x=0$ is an asymptotically stable equilibrium for $x^+=f(x)$. Then the ROA  of $x^+=f(x)$ is given by
\[
\mathcal{A}_0=\{
x_0\colon  \lim_{k\to\infty} 
s_k(x_0) =0
\}
\]
where $s_k(x_0)$ is the solution to $x^+=f(x)$ at time $k\ge k_0$ from the initial condition $x_0$. \qeds
\end{definition}

\begin{definition}
\cite[Definition 13.4]{haddad2011nonlinear}
A set $\mathcal{M}\subset \mathbb{R}^n$ is a positively invariant set for $x^+=f(x)$ if $s_k(\mathcal{M})\subseteq \mathcal{M}$ for all $k\ge k_0$, where $s_k(\mathcal{M})=\{s_k(x_0)\colon x_0 \in \mathcal{M}\}$. \qeds
\end{definition}

Recall the Lyapunov difference 
\[\ba{rl}
& V(w^+, \xi^+)-V(w,\xi)\\
=& 
\left(M \begin{bmatrix}
w\\
\xi
\end{bmatrix}+N Q(w,\xi)
\right)^\top \mathcal{P}_1^{-1} 
\left(M \begin{bmatrix}
w\\
\xi
\end{bmatrix}+N Q(w,\xi)
\right)
\\
&
-
\begin{bmatrix}
w\\
\xi
\end{bmatrix}^\top
\mathcal{P}_1^{-1}
\begin{bmatrix}
w\\
\xi
\end{bmatrix}=: \mathcal{W}(w,\xi)
\ea\]
with $M,N$ as in \eqref{M-N}.
\begin{cor}\label{corol1}
Consider the same setting as Proposition \ref{prop:approx}. Let\footnote{Although not indicated explicitly, $\mathcal{V}$ is a subset of the domain of definition of $V(w,\xi)$.
}
 $\mathcal{V}:=\{(w,\xi) \colon \mathcal{W}(w,\xi) <0\}$. Any sublevel set $\mathcal{R}_\gamma=\{(w,\xi) \colon V(w,\xi) \le \gamma\}$ contained in $\mathcal{V}\cup \{0\}$ is positively invariant for system 
\eqref{nonl-eq-extended-actual-closed-loop} and defines an estimate of the  
ROA of system \eqref{nonl-eq-extended-actual-closed-loop}.
\qeds
\end{cor}

\smallskip

As the function $\mathcal{W}(w,\xi)$ is known from the data, the estimate of the ROA $\mathcal{R}_\gamma$ is computable. 

\section{Main result}\label{sec:convergence-true-state}

To draw conclusions on the convergence of system \eqref{nonl}, we first observe that 
the dynamical controller \eqref{chain-integrators-with-u} uses its own state $\xi$ and the state $w$ to generate the control action $u=\kappa Z(\xi,w)$. At time $k$ the state $w(k)$ contains the past $N$ output measurements from the process \eqref{nonl}, from which we only measure $y(k)$. To make the past measurements in $w(k)$  available to the controller, we extend it with the dynamics
\be\label{chain-integrators-with-y}
\eta^+ = A_c \eta + B_c y
\ee
Then, for any $k_0\in \mathbb{Z}$ and any $\eta(k_0)\in \mathbb{R}^N$, we have that $\eta(k)=y_{[k-N,k-1]}=w(k)$ for all $k\ge k_0+N$, that is, independently of the initialization of \eqref{chain-integrators-with-y}, its state $\eta(k)$ provides the vector $w(k)$ of the past output measurements from time $N$ onward. Similarly, for any $\xi(k_0)\in \mathbb{R}^N$, system \eqref{chain-integrators-with-u} is such that  $\xi(k)=u_{[k-N,k-1]}$ for all $k\ge k_0+N$. See \cite{messina2005discrete} for the same structure of the controller
\eqref{chain-integrators-with-u}, \eqref{chain-integrators-with-y}.

\medskip

\begin{rem} System \eqref{chain-integrators-with-y} is the so-called deadbeat observer, since for $k\ge k_0+N$, the mapping $\psi(\eta(k),\xi(k))$ would return $x(k)$. If both $\psi$ {\em and} a state-feedback stabilizer for system \eqref{nonl} were known, one could obtain a dynamic output feedback controller for the system \eqref{nonl}. Here we are interested to the case in which this knowledge is not available and we design a dynamic output feedback controller under a suitable assumption on the nonlinearity $\tilde h$ (Assumption \ref{asspt-dictionary}). 
\qedp
\end{rem}

\smallskip

The following statement transfers the result obtained for the system 
\eqref{nonl-eq-extended-actual-closed-loop} to the actual closed-loop system that includes the process \eqref{nonl}.
\begin{prop}\label{prop:approx-cancell-via-output-feedback}
Let Assumptions \ref{asspt-uo-on-set}, \ref{asspt-dictionary} and \ref{asspt:dataset-in-the-right-domain} hold. 
Consider the  SDP \eqref{eq:2SDP}, assume that it is feasible and let condition \eqref{eq:limitQx} hold. 
For any $(x_0, \xi_0, \eta_0)\in \mathcal{X}\times  \mathbb{R}^N\times  \mathbb{R}^N$ for which there exists $v=(v_{[0,N-2]},v_{N-1})\in \mathcal{U}^{N}$
such that   $(\Phi_N(x_0,v_{[0,N-2]}),v) \in \mathcal{R}_\gamma$, the solution of the system \eqref{nonl} in closed-loop with the time-varying controller comprised by \eqref{chain-integrators-with-u}, \eqref{chain-integrators-with-y}
and  
\be\label{u-tv}
u(k)= 
\left\{
\ba{ll}
v_{k-k_0} & k_0\le k\le k_0+N-1\\
\kappa Z(\eta(k),\xi(k)) & k\ge k_0+N
\ea
\right.
\ee
that starts from $(x_0, \xi_0, \eta_0)$, asymptotically converges to the origin.  \qeds
\end{prop}

{\it Proof.} First note that, by definition of the mapping $\Phi_N$ and since $f(0,0)=0$ and $h(0)=0$, each entry of $\Phi_N$ is a continuous function of its arguments which is zero when these are zero, hence there exists a neighbourhood of the origin $(x,v)=(0,0)$ such that any point $(x,v)$ in the neighbourhood satisfies $(\Phi_N(x,v_{[0, N-2]}),v) \in \mathcal{R}_\gamma$. 

By definition of the mapping $\Phi_N$ in Assumption \ref{asspt-uo-on-set} and \eqref{u-tv}, $\Phi_N(x_0,v_{[0,N-2]})=
y_{[k_0,k_0+N-1]}$, where $y$ denotes the output response of the closed-loop system from the initial condition $(x_0, \xi_0, \eta_0)$.  

 By the dynamics of the controller \eqref{chain-integrators-with-u}, \eqref{chain-integrators-with-y}, we have $\eta(k)=y_{[k-N, k-1]}$, $\xi(k)=u_{[k-N,k-1]}$ for all $k\ge k_0+N$ and $(\eta(k_0+N), \xi(k_0+N))=(\Phi_N(x,v_{[0,N-2]}),v)\in \mathcal{R}_\gamma$. Hence, by Lemma \ref{lem-equiv}, the solution of \eqref{chain-integrators-with-u}, \eqref{chain-integrators-with-y} are the same as those of
 system 
 \eqref{nonl-eq-extended-actual-closed-loop} intialized at $(w(k_0+N), \xi(k_0+N))=(y_{[k_0,k_0+N-1]}, u_{[k_0,k_0+N-1]})$. As $(\eta(k_0+N), \xi(k_0+N))\in \mathcal{R}_\gamma$, by Proposition \ref{prop:approx} and Corollary \ref{corol1}, $(\eta(k), \xi(k))$ converges to the origin. By Lemma \ref{lem-equiv},  for all $k\ge k_0+N$, $x(k)= \psi(\eta(k),\xi(k))$, which implies convergence of $x(k)$ to the origin by continuity of $\psi$. \qedp

\medskip

The particular form of $u(k)$ in \eqref{u-tv} is due to the fact that, during the first $N$-steps, the controller state does not provide an accurate value of the past input-output measurements of the system, hence the choice to apply an open-loop input sequence. After $N$ time steps, when such past measurements become available through the controller states $\eta(k), \xi(k)$,  $u(k)$ is set to the feedback $\kappa Z(\eta(k),\xi(k))$.

We also remark that in the result above if the initial condition $x_0$ is sufficiently close to the origin and the initial sequence of control values $v_0,\ldots,v_{n-2},v_{n-1}$ does not drive the output response of \eqref{nonl} outside the set $\mathcal{R}_\gamma$, then the designed controller \eqref{u-tv} steers the state of the overall closed-loop system to the origin. Note that $\mathcal{R}_\gamma$ is known thanks to Corollary \ref{corol1}, hence the designer can check whether the initial control sequence and the corresponding measured output response are in $\mathcal{R}_\gamma$. For the design of the initial control sequence, the designer could take advantage of some expert knowledge.


%

\begin{rem}
\emph{(Prior on input/output measurements)} The controller is designed under the assumption that the input/output measurements collected during the experiment range over some specified sets -- see Assumption \ref{asspt:dataset-in-the-right-domain} -- where the measurements provide meaningful information about the system's internal state.  
These sets are not known, hence, the feature that the evolution of the system during the experiments remains in the sets of interest must be considered as one of the priors under which the design is possible.
\end{rem}

\section{Numerical example}
We continue with Example \ref{example} 
and consider the equations 
\eqref{eq:sys_example_pendulum} with 
output $y=x_1$. The system parameters are 
$T_s = 0.1$, $m = 1, \ell = 1, g = 9.8$ and $\mu =0.01$. The problem is to learn a  controller for \eqref{eq:sys_example_pendulum} from input-output data that renders the origin of the closed-loop system locally asymptotically stable.
\par

Following \cite[Example 5]{2021cancellation}, we choose 
\[
Z(w,\xi)= 
\begin{bmatrix}
w\\
\xi\\
\sin{w_1} - w_1 \\
\xi_1 \cos{w_1} - \xi_1
\end{bmatrix}
\]
and note that Assumption \ref{asspt-dictionary} and \eqref{eq:limitQx} hold. 

\begin{figure}[htb]
    \centering    \includegraphics[scale=0.6]{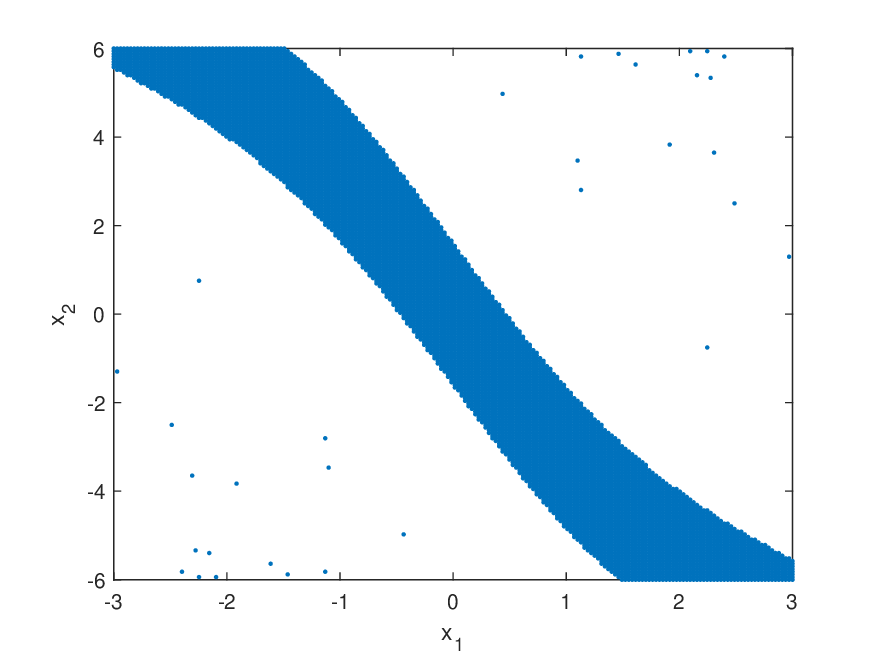}\caption{The blue area represents the estimate of the ROA of system \eqref{eq:sys_example_pendulum} in closed-loop with the controller \eqref{chain-integrators-with-u}, \eqref{chain-integrators-with-y}, \eqref{u-tv}, where 
    $v_{k-k_0}=0$ for  $k_0\le k\le k_0+N-1$ and 
    $\kappa$ is given in \eqref{kappa.example}.}
    \label{wcoorex1}
\end{figure}

We collect data by  running $T=7$, $N=2$-long experiments  with input uniformly distributed in $[-0.5, 0.5]$ and with an initial state in $[-0.5, 0.5]^2$. For each experiment $j = 0, 1, \ldots, T-1$, we collect the samples $\{ u^j(k), y^j(k) \}_{k=0}^2$. Then we construct data matrices $Y_1, Y_0, V_1, V_0, U_0, Q_0$, as detailed in Remark \ref{rem.mult-experiments}. The program \eqref{eq:2SDP}  is feasible and we obtain the controller gain with 
\be\label{kappa.example}
\kappa = \begin{bmatrix} 52.4412 & - 76.1179 & -0.5782 & - 0.4467 & 0 & 0   \end{bmatrix}
\ee
using the YALMIP toolbox \cite{lofberg2004yalmip}, MOSEK solver \cite{aps2019mosek}. To assess the effectiveness of the designed controller, instead of computing $\mathcal{R}_\gamma$, which for this example provides a  conservative estimate of the ROA, we depict in Fig.~\ref{wcoorex1} the set of initial conditions $x_0$ for which, choosing $v_{k-k_0}=0$ for  $k_0\le k\le k_0+N-1$ in \eqref{u-tv}, the state $(x(k),\eta(k), \xi(k))$ converges to zero. Note that the choice of $\eta_0, \xi_0$ is inessential. 
The  set is obtained by letting the closed-loop system evolve for $200$ time steps and then checking whether or not the 
norm 
$\|(x(k),\eta(k), \xi(k))\|_\infty$ is smaller than $10^{-6}$ on the interval $195\le k\le 200$. 

\section{CONCLUSIONS}

We have examined a design of dynamic output feedback controllers for nonlinear systems from input/output data. 
The uniform observability property of the system, a prior in the approach, is instrumental to define a new set of coordinates, from which a data-driven ``state"-feedback design can be conducted. The result is local and the size of the region of attraction is limited by the free evolution of the system during the first $N$ steps during which the dead-beat observer reconstructs the past input/output values that feed the controller.  The design and analysis have been carried out in the favourable setting in which measurements are noise-free and the nonlinearities can be expressed via a dictionary of known functions. 
Regarding the future work, besides going beyond the favourable setting, we would like to explore 
either a more sophisticated observer design or a different data-driven control design method. An option is to express the function $\psi$ via a dictionary of functions, perform a data-driven design of an observer and follow a certainty equivalence principle in the analysis of the closed-loop system.

\addtolength{\textheight}{-12cm}   



\section*{APPENDIX}

\subsection{Proof of Lemma \ref{lem-equiv}}\label{app:proof-lemma-1}
For any time $k$,  collect the past $N$ output samples generated by the system \eqref{nonl} in the vector 
$y_{[k-N,k-1]}$. 
The vectors $y_{[k-N,k-1]}, y_{[k-N+1,k]}$ at two successive time instants are related by
\be\label{dyn-Y}
y_{[k-N+1,k]} = \begin{bmatrix}y(k-N+1)\\ y(k-N+2)\\ \vdots \\ y(k-1)\\ y(k)\end{bmatrix}
=
A_c y_{[k-N,k-1]} + B_c y(k)
\ee
By the dynamics \eqref{nonl} and the definitions \eqref{F}, the state $x(k)$ at time $k$ 
is given by
\be\label{state.k}
x(k)=F^N(x(k-N), u_{[k-N,k-1]})
\ee
the output $y(k)$ at time $k$ is given by
\[
y(k)= h\circ F^N(x(k-N), u_{[k-N,k-1]})
\]
and, by the definition \eqref{Phi_N},
\be\label{mapping-returning-x}
y_{[k-N,k-1]}= \Phi_N(x(k-N), u_{[k-N,k-2]}).
\ee
By Assumption \ref{asspt-uo-on-set} and the hypothesis that $x(k-N)\in \mathcal{X}$ and $u_{[k-N, k-1]}\in \mathcal{U}^N$ for all $k\in \mathbb{Z}_{\ge k_0}$, the mapping \eqref{mapping-returning-x} is invertible and returns 
\[
x(k-N) = \Psi_N(y_{[k-N,k-1]}, u_{[k-N, k-2]}).
\] 
Hence, the state $x(k)$ in \eqref{state.k} can be expressed as a mapping of the past input and output samples
\[\ba{rl}
x(k)=& F^N(\Psi_N(y_{[k-N,k-1]}, u_{[k-N, k-2]}), u_{[k-N, k-1]}) \\
=& \psi(y_{[k-N,k-1]}, u_{[k-N, k-1]})
\ea
\]
and similarly for the output
\[
\ba{rl}
y(k)=& h\circ F^N(\Psi_N(y_{[k-N,k-1]}, u_{[k-N, k-2]}), u_{[k-N, k-1]})\\
=& \tilde h(y_{[k-N,k-1]}, u_{[k-N, k-1]})
\ea
\] 
If $y(k)$ is replaced in \eqref{dyn-Y}, then
 \[
y_{[k-N+1,k]}= \tilde f(y_{[k-N,k-1]},u_{[k-N, k-1]}),
\]
by definition of $\tilde f$ in \eqref{mappings}. 

By the choice of the input 
$v(k)=u_{[k-N, k-1]}$, for all $k\in \mathbb{Z}_{\ge k_0}$  
and of the initial condition $w(k_0)=y_{[k_0-N, k_0-1]}$, we have that $w(k)=y_{[k-N,k-1]}$ for all $k\in \mathbb{Z}_{\ge k_0}$, 
and this ends the proof. 
\qedp

%
%
%


\bibliographystyle{IEEEtran}

\bibliography{refs-root}

\end{document}